\documentclass
[%
reprint,
showpacs,preprintnumbers,
amsmath,amssymb,
aps,
pre,
floatfix,
longbibliography
]{revtex4-1}

\usepackage{booktabs}
\usepackage{array}
\newcolumntype{L}[1]{>{\raggedleft\arraybackslash}p{#1}}
\AtBeginDocument{
	\heavyrulewidth=.08em
	\lightrulewidth=.05em
	\cmidrulewidth=.03em
	\belowrulesep=.65ex
	\belowbottomsep=0pt
	\aboverulesep=.4ex
	\abovetopsep=0pt
	\cmidrulesep=\doublerulesep
	\cmidrulekern=.5em
	\defaultaddspace=.5em
}

\usepackage{color}
\usepackage{xcolor}
\usepackage{graphicx}
\usepackage{dcolumn}
\usepackage{bm}
\usepackage{cancel}
\usepackage[utf8]{inputenc}

\begin{document}

\title{Dependence on the thermodynamic state of self-diffusion of pseudo hard-spheres}


\author{L. Marchioni}
\author{M. A. Di Muro}  
\author{M. Hoyuelos}
\email{hoyuelos@mdp.edu.ar}

\affiliation{Instituto de Investigaciones F\'isicas de Mar del Plata (IFIMAR -- CONICET), Departamento de F\'isica, Facultad de Ciencias Exactas y Naturales,
	Universidad Nacional de Mar del Plata, De\'an Funes 3350, 7600 Mar del Plata, Argentina}

\date{\today}

\begin{abstract}
Self-diffusion, $D$, in a system of particles that interact with a pseudo hard sphere potential is analyzed. Coupling with a solvent is represented by a Langevin thermostat, characterized by the damping time $t_d$. The hypotheses that $D=D_0 \varphi$ is proposed, where $D_0$ is the small concentration diffusivity and $\varphi$ is a thermodynamic function that represents the effects of interactions as concentration is increased. Molecular dynamics simulations show that different values of the noise intensity modify $D_0$ but do not modify $\varphi$. This result is consistent with the assumption that $\varphi$ is a thermodynamic function, since the thermodynamic state is not modified by the presence of damping and noise.
\end{abstract}


\maketitle

\section{Introduction}

The Chapman-Enskog transport theory \cite{chapman} predicts self-diffusivity of a general system of interacting particles at small concentration; it reduces to the Boltzmann theory for dilute gases when particles are hard spheres. Several approximate theories were proposed to extend the description to moderate and large concentrations, such as free volume \cite{dymond,hildebrand,batschinski,doolittle,cohen2,turnbull,macedo} or excess entropy theories \cite{rosenfeld,rosenfeld2}. They were successfully applied to fluid models and real substances for the description of transport properties, usually requiring around 2--4 adjustable parameters.

Nevertheless, a complete theoretical understanding of self-diffusion of hard spheres remains an open problem. Despite its simplicity, the thermodynamic and transport properties of the hard sphere model do not yet have exact solutions in the whole range of concentration. Accurate mathematical representations of self-diffusivity for hard spheres are in general obtained from interpolation of molecular dynamics results (see Sect.\ 9.4 in \cite{silva} for a review). See Pieprzyk \textit{et al.} \cite{pieprzyk2} for more recent numerical simulations.

The purpose ot this paper is to elucidate qualitative aspects of self-diffusion of hard spheres in three dimensions assuming, as a starting point, the following hypotheses: the self-diffusion coefficient, $D$, can be written as the product of two terms, one including the dependence on factors such as mass of particles, size of particles or mean velocity (or temperature), and the other corresponding to macroscopic or thermodynamic aspects associated to the presence of interactions. It is assumed that the effects of interactions, that manifest themselves when concentration is increased, are represented by the second term. For brevity, we refer to the ``self-diffusion coefficient'' as ``diffusivity''. We identify the first term with the diffusivity at small concentration, $D_0$, so that
\begin{equation}\label{e.D}
D = D_0 \varphi
\end{equation}
where $\varphi$ is the factor representing the effect interactions as concentration is increased. According to the hypotheses, only the information of interactions at the thermodynamic level is needed to describe diffusivity in the whole concentration range.

The method proposed to check this hypothesis is as follows. Noise corresponding to a Langevin thermostat is added; the system can be seen as a colloidal suspension of hard spheres where hydrodynamic forces are neglected \cite{dhont}. The noise modifies $D_0$, but it does not modify the thermodynamic state, so that $\varphi$ should remain unchanged. Molecular dynamics simulations are carried out using the continuous pseudo hard-sphere potential proposed in Ref.\ \cite{jover}, that accurately reproduces results of the hard-sphere potential; see Sec.\ \ref{s.potential}. Simulations are preformed, using LAMMPS software \cite{plimpton}, to verify that the equation of state does not change with the noise intensity (represented by the damping time $t_d$); see Sec.\ \ref{s.eos}. This means that the presence of noise does not modify the thermodynamic state and, therefore, it does not modify $\varphi$. In Sec.\ \ref{s.d0}, the form of diffusivity at small concentration is determined as a combination of Langevin and Boltzmann diffusion coefficients. Numerical results of $D/D_0$ against concentration for different values of the noise intensity are presented in Sec.\ \ref{s.numerical}. The results show that, as expected, $\varphi$ does not change for different noise intensities. Summary and conclusions are presented in Sec.\ \ref{s.conclusions}.

\section{Pseudo hard-sphere potential}
\label{s.potential}

The Mie potential, as the Lennard-Jones potential, is repulsive at short radial distance $r$ and has an attractive well of energy $\epsilon$ at intermediate distances. It generalizes the Lennard-Jones potential by considering exponents $\lambda_r$ and $\lambda_a$ of the repulsive and attractive terms:
\begin{equation}\label{e.Mie}
u_\text{Mie}(r) = \frac{\lambda_r}{\lambda_r-\lambda_a}\left( \frac{\lambda_r}{\lambda_a} \right)^{\frac{\lambda_a}{\lambda_r-\lambda_a}} \epsilon \left[ \left( \frac{\sigma}{r} \right)^{\lambda_r} - \left( \frac{\sigma}{r} \right)^{\lambda_a} \right].
\end{equation}
where the size parameter, $\sigma$, is related to the diameter of the spherically symmetric  particles. Weeks, Chandler and Andersen \cite{weeks} proposed a cut and shifted version of the Lennard-Jones potential in order to consider a purely repulsive core. The same procedure applied to the Mie potential with $\lambda_r=50$ and $\lambda_a=49$ results in:
\begin{equation}\label{e.pseudo}
u(r) = \left\{ \begin{array}{cc}
50 (\frac{50}{49})^{49} \epsilon \left[(\frac{\sigma}{r})^{50} - (\frac{\sigma}{r})^{49}\right] + \epsilon & \ \ \ r < \sigma \frac{50}{49} \\[\medskipamount]
0 & \ \ \ r \ge \sigma \frac{50}{49}
\end{array}
  \right. .
\end{equation}
This is the continuous potential used in Ref.\ \cite{jover} to reproduce results of hard spheres. They have shown that the correspondence with hard spheres is fulfilled for a reduced temperature $T^* = k_B T/\epsilon = 1.5$, where $T$ is temperature and $k_B$ is Boltzmann's constant. The particle mass, $m$, the energy $\epsilon$ and the size $\sigma$ are combined to cancel units in reduced quantities.

\section{Equation of state}
\label{s.eos}

Carnahan and Starling \cite{carnahan} obtained an approximate equation of state (EOS) for the hard sphere fluid that is widely used due to its simplicity and accuracy. It is an expression for the compressibility factor, defined as
\begin{equation}\label{e.Z}
Z = \frac{PV}{N k_B T} = \frac{P^*}{\rho^* T^*}
\end{equation}
where $P$ is the pressure, $V$ is the volume and $N$ is the number of particles; variables with asterisks are dimensionless reduced quantities defined as $\rho^* = \sigma^3 N/V$ and $P^* = P \sigma^3/\epsilon$. The Carnahan and Starling EOS is
\begin{equation}\label{e.eos}
Z = \frac{1 + \eta + \eta^2 - \eta^3}{(1-\eta)^3},
\end{equation}
where $\eta = \frac{\pi \sigma^3}{6}\frac{N}{V}=\rho^* \pi/6$ is the packing fraction; the equation holds for $\eta< 0.55$. The hard sphere EOS does not depend on temperature. Instead, the EOS for the pseudo hard sphere potential \eqref{e.pseudo} depends on temperature and, in order to reproduce the hard sphere behavior, temperature has to be fixed at $T^*=1.5$, as mentioned in the previous section.

We consider that the hard sphere system is immersed in a background solvent modeled by a Langevin thermostat. Two forces are introduced by the thermostat: a friction force given by $-m \mathbf{v}/t_d$, where $\mathbf{v}$ is the particle's velocity and $t_d$ is the damping time, and a stochastic force represented by white noise of intensity $2 m k_B T /t_d$. Since temperature is fixed, both forces are determined by the value of the damping time (equal to the inverse of the friction coefficient). The reduced damping time is $t^*_d = t_d\, \sigma^{-1}\sqrt{\epsilon/m}$.

\begin{figure}
		\begin{center}
		\includegraphics[width=9cm]{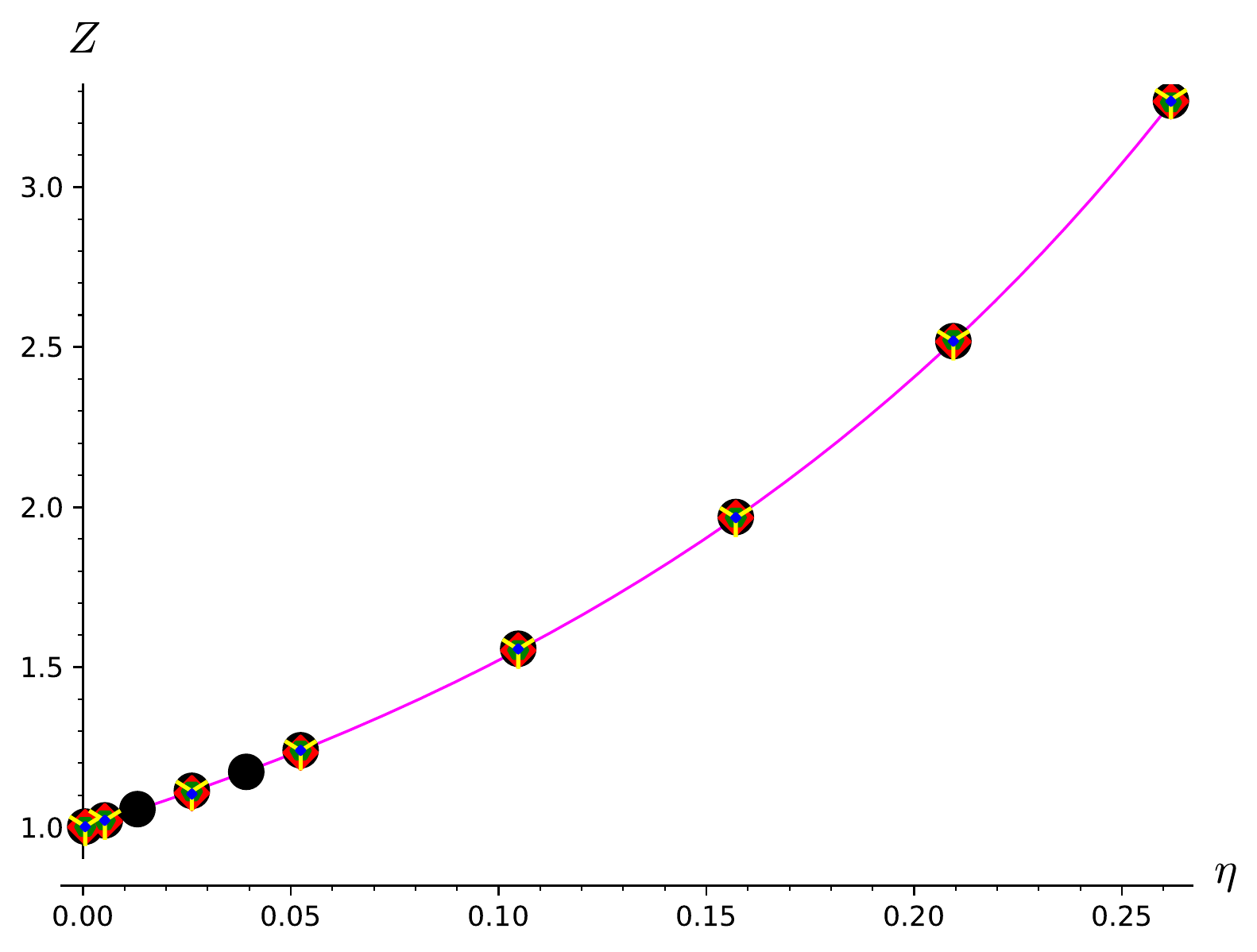}
	\end{center}
	\caption{Compressibility factor $Z$ against packing fraction for different values of the damping time $t_d^*$: $0.01$, $0.1$, 1 and 10. The curve corresponds to the Carnahan and Starling EOS, Eq.\ \eqref{e.eos}.}\label{f.eos}
\end{figure}

The pressure, and the compressibility factor, were numerically calculated for concentrations in the fluid range and for different values of $t_d^*$, see Fig.\ \ref{f.eos}. The figure shows that the EOS is not modified by the presence of noise, meaning that the value of $\varphi$ in \eqref{e.D} should not change with $t_d^*$. 

\section{Diffusivity at small concentration}
\label{s.d0}

In order to numerically determine the behavior of $\varphi$ we need to calculate the ratio $D/D_0$. The diffusivity at small concentration, $D_0$, is well known for the two extreme values of the damping time. If $t_a^*\rightarrow \infty$, noise and damping are absent and, according to Boltzmann's theory, the diffusivity at small concentration is given by
\begin{equation}\label{e.dboltz}
D_B = \frac{3}{8 \rho \sigma^2}\sqrt{\frac{k_B T}{\pi m}}.
\end{equation}
The reduced diffusivity is $D^* = D \sigma^{-1}\sqrt{m/\epsilon}$, then
\begin{equation}\label{e.dboltza}
D_B^* = \frac{\sqrt{\pi T^*}}{16\, \eta}.
\end{equation}

In the other extreme, for small $t_d^*$, noise and damping dominate the hard sphere behavior and, according to Langevin's theory, the diffusivity is
\begin{equation}\label{e.dlang}
D_L = \frac{k_B T t_d}{m},
\end{equation}
or
\begin{equation}\label{e.dlanga}
D_L^*= T^* t_d^*.
\end{equation}

For intermediate values of $t_d^*$, $D_0$ is given by a combination of $D_L$ and $D_B$. The inverse of the diffusivity is the resistance to the particle current. When collisions between hard spheres and interactions with the background solvent are both relevant, the associated resistances are added. The small concentration diffusivity is
\begin{equation}\label{e.D0}
\frac{1}{D_0^*} = \frac{1}{D_B^*} + \frac{1}{D_L^*},
\end{equation}
or
\begin{equation}\label{e.D02}
D_0^* = \frac{T^* t_d^*}{16 \eta t_d^* \sqrt{T^*/\pi} + 1}.
\end{equation}
For the two extreme values of the damping time we have $D_0^* \simeq D_L^*$ (small $t_d^*$) and $D_0^* \simeq D_B^*$ (large $t_d^*$).

\begin{figure}
	\begin{center}
		\includegraphics[width=8cm]{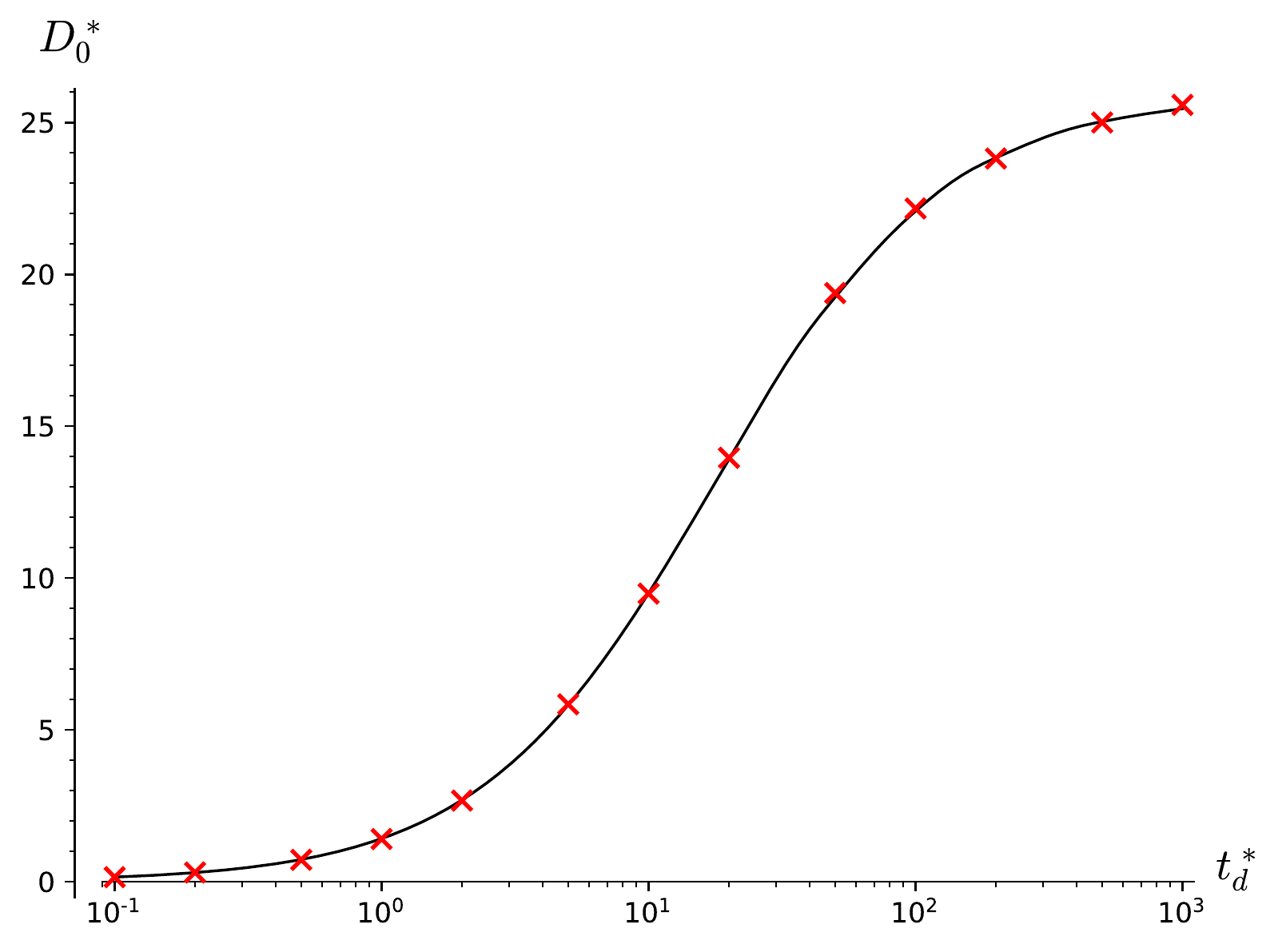}
	\end{center}
	\caption{Reduced diffusivity at small concentration, $D_0^*$, against dumping time, $t_d^*$, for $\eta=0.0052$. Crosses are numerical results and the curve corresponds to Eq.\ \eqref{e.D0}.}\label{f.D0}
\end{figure}

The validity of Eq.\ \eqref{e.D0} was checked with numerical simulations at small concentration. Fig.\ \ref{f.D0} shows $D_0^*$ against the damping time, $t_d^*$, for $\eta=0.0052$. Numerical values of $D_0^*$ coincide with Eq.\ \eqref{e.D0}.

\section{Numerical results of $D/D_0$}
\label{s.numerical}

According to our hypotheses, the diffusivity $D$ is proportional to $D_0$ for the whole concentration range, see Eq.\ \eqref{e.D}. The correction factor $\varphi$ is equal to 1 for small concentration. As concentration  increases, diffusion decreases and $\varphi$ decays to zero due to  clogging of the system. The point that we wish to verify is that this decay, characterized by $\varphi$, depends on the thermodynamic state, that is unchanged by the presence of noise. Therefore, $\varphi$ should not depend on $t_d^*$.

\begin{figure}
		\begin{center}
		\includegraphics[width=8cm]{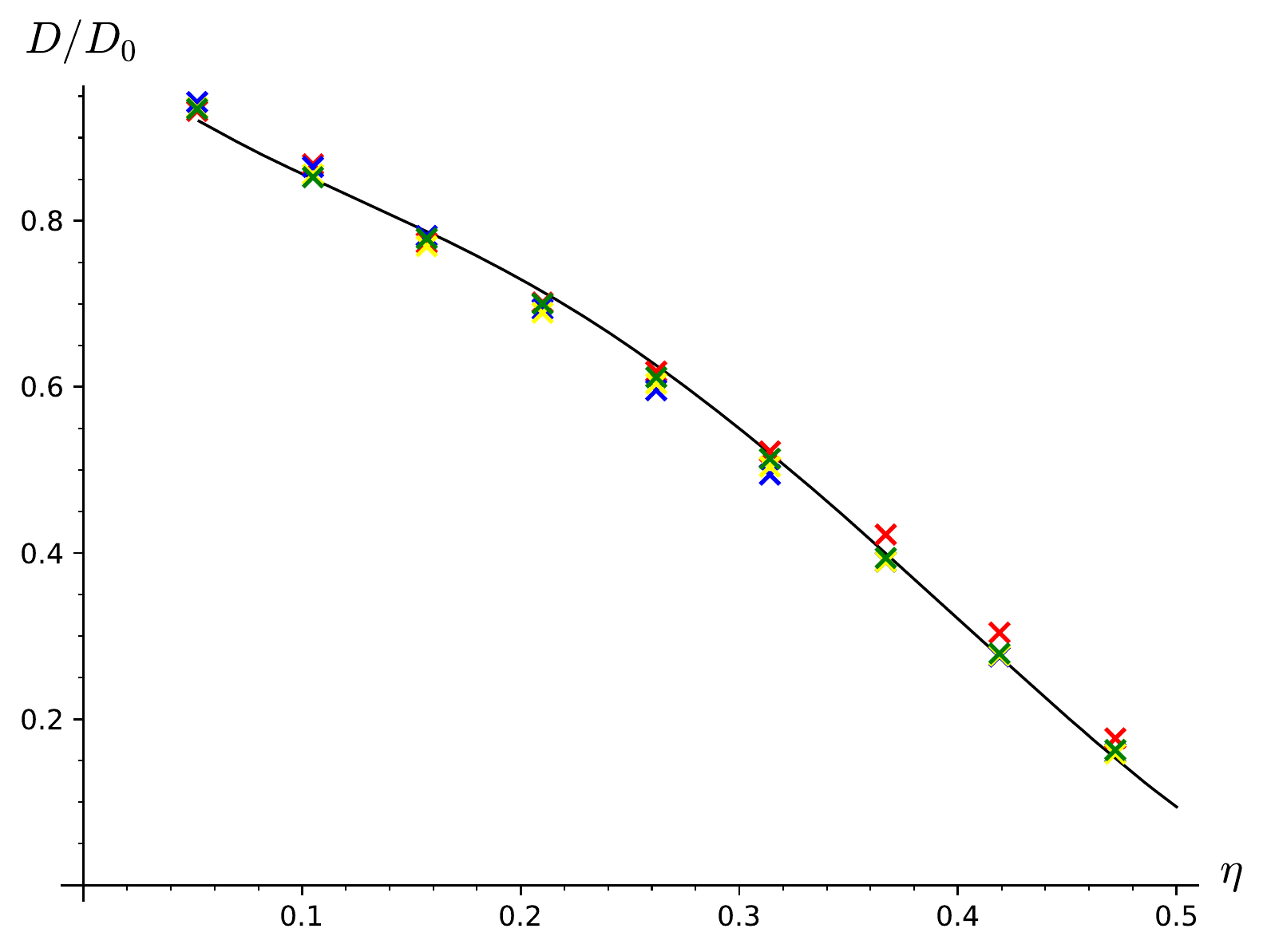}
	\end{center}
	\caption{$D/D_0$ against packing fraction, $\eta$, for different values of the damping time $t_d^*$, red for $0.1$, blue for $1$, yellow for $5$ and green for $10$. The superimposed curve corresponds to numerical results of Pieprzyk \textit{et al.} \cite{pieprzyk2} for hard spheres. Parameters of the simulations: Number of samples: $10$, maximun time (after thermalization): $25$, time step: $0.001$ and number of particles por sample: $500000$. }\label{f.phi}
\end{figure}

Numerical values of $D$ against concentration were obtained for different values of the damping time. The method to calculate diffusivity was the integration of the velocity auto-correlation (Green-Kubo formula). Fig.\ \ref{f.phi} shows values of $D/D_0$ against packing fraction. It can be seen that the behavior is not modified by changing noise and damping; values of $D/D_0$ for different $t_d^*$ coincide within numerical fluctuations. Numerical results of Pieprzyk \textit{et al.} \cite{pieprzyk2} for hard spheres (without noise) are also plotted for comparison.

\section{Conclusions}
\label{s.conclusions}

Self-diffusion in a system of particles that interact through a pseudo hard-sphere potential is analyzed. Particles are in a solvent whose effects are represented by a Langevin thermostat. The thermodynamic state is independent of the coupling with the thermostat, given by the damping time $t_d^*$. Numerical simulations were performed to check that the equation of state is independent of $t_d^*$ and also to verify that the small concentration diffusivity, $D_0$, can be written as a combination of the diffusivities from Boltzmann and Langevin theories.

It is proposed, as an hypotheses, that the self-diffusion coefficient is proportional to $D_0$ and that the proportionality factor, $\varphi$, is a thermodynamic function; see Eq.\ \eqref{e.D}. In the limit of small concentration, $\varphi$ goes to one. Numerical simulations are consistent with this hypotheses. Values obtained of $D/D_0$ against the packing fraction are independent of the damping time, $t_d^*$. This result is compatible with the assumption that $\varphi$ is a thermodynamic function.

Knowing that $\varphi$ is a thermodynamic function is a useful guide for the development of a theory for diffusion of hard spheres. As mentioned in the introduction, the concentration dependence of $\varphi$ is an open problem. Factor $\varphi$ contains information about the influence of interactions on diffusion, but microscopic details of the interaction potential are not needed. We can expect that the interaction information contained in the excess chemical potential is important for the determination of $\varphi$ for hard spheres. The hypotheses still has to be verified for other interaction potentials.

\bibliographystyle{elsarticle-num}
\bibliography{noise.bib}

\begin{thebibliography}{10}
\expandafter\ifx\csname url\endcsname\relax
  \def\url#1{\texttt{#1}}\fi
\expandafter\ifx\csname urlprefix\endcsname\relax\def\urlprefix{URL }\fi
\expandafter\ifx\csname href\endcsname\relax
  \def\href#1#2{#2} \def\path#1{#1}\fi

\bibitem{chapman}
S.~Chapman, T.~G. Cowling, The Mathematical Theory of Non-Uniform Gases, 3rd
  Edition, Cambridge University Press, 1970.

\bibitem{dymond}
J.~H. Dymond, Corrected {E}nskog theory and the transport coefficients of
  liquids, J. Chem. Phys. 60 (1974) 969.

\bibitem{hildebrand}
J.~H. Hildebrand, Motions of molecules in liquids: Viscosity and diffusivity,
  Science 174 (1971) 490.

\bibitem{batschinski}
A.~J. Batschinski, Investigations of internal friction of fluids, Z. Phys.
  Chem. 84 (1913) 643.

\bibitem{doolittle}
A.~K. Doolittle, Studies in {N}ewtonian flow. {II}. {T}he dependence of the
  viscosity of liquids on free‐space, J. Appl. Phys. 22 (1951) 1471.

\bibitem{cohen2}
M.~H. Cohen, D.~Turnbull, Molecular transport in liquids and glasses, J. Chem.
  Phys. 31 (1959) 1164.

\bibitem{turnbull}
D.~Turnbull, M.~H. Cohen, On the free‐volume model of the liquid‐glass
  transition, J. Chem. Phys. 52 (1970) 3038.

\bibitem{macedo}
P.~B. Macedo, T.~A. Litovitz, On the relative roles of free volume and
  activation energy in the viscosity of liquids, J. Chem. Phys. 42 (1965) 245.

\bibitem{rosenfeld}
Y.~Rosenfeld, Relation between the transport coefficients and the internal
  entropy of simple systems, Phys. Rev. A 15 (1977) 2545.

\bibitem{rosenfeld2}
Y.~Rosenfeld, Comments on the transport coefficients of dense hard core
  systems, Chem. Phys. Lett. 48 (1977) 467.

\bibitem{silva}
C.~M. Silva, H.~Liu, Modelling of transport properties of hard sphere fluids
  and related systems, and its applications, in: A.~Mulero (Ed.), Theory and
  Simulation of Hard-Sphere Fluids and Related Systems, Springer, 2008, p. 383.

\bibitem{pieprzyk2}
S.~Pieprzyk, M.~N. Bannerman, A.~C. Bra\'{n}ka, M.~Chudak, D.~M. Heyes,
  Thermodynamic and dynamical properties of the hard sphere system revisited by
  molecular dynamics simulation, Phys. Chem. Chem. Phys. 21 (2019) 6886.

\bibitem{dhont}
J.~K.~G. Dhont, An Introduction to Dynamics of Colloids, Elsevier, Amsterdam,
  1996.

\bibitem{jover}
J.~Jover, A.~J. Haslam, A.~Galindo, G.~Jackson, E.~A. M{\"u}ller, Pseudo
  hard-sphere potential for use in continuous molecular-dynamics simulation of
  spherical and chain molecules, J. Chem. Phys. 137 (2012) 144505.

\bibitem{plimpton}
S.~Plimpton, Fast parallel algorithms for short-range molecular dynamics,
  Journal of Computational Physics 117~(1) (1995) 1--19, see
  http://lammps.sandia.gov.

\bibitem{weeks}
J.~Weeks, D.~Chandler, H.~Andersen, The role of repulsive forces in determining
  the equilibrium structure of simple liquids, J. Chem. Phys. 54 (1971) 5237.

\bibitem{carnahan}
N.~F. Carnahan, K.~E. Starling, Equation of state for nonattracting rigid
  spheres, J. Chem. Phys. 51 (1969) 635.

\end{thebibliography}

\end{document}